\documentclass[12pt]{article}

\def\pd{\partial_{\mu}}
\def\pa{\partial}
\def\pu{\partial^{\mu}}

\def\p{\phi}
\def\k{\psi}

\def\l{\lambda}
\def\m{\mu}
\def\n{\nu}
\def\e{\epsilon}
\def\s{\sigma}

\def\g{\gamma}

\def\a{\alpha}
\def\b{\beta}
\def\r{\rho}

\def\vt{\vartheta}

\def\F{{\cal F}}

\def\A{{\cal A}}
\def\be{\begin{equation}}
\def\ee{\end{equation}}
\def\ba{\begin{eqnarray}}
\def\ea{\end{eqnarray}}
\def\z{{ }^{*}\hspace{-.5mm}}
\def\hn{\hat N}
\def\h{\hat}

\def\f0{f_0}
\begin{document}
\thispagestyle{empty}
\vspace{3cm}
\rightline{SINP-2002-7/691}
\vspace{4cm}
\centerline{\large\bf  Dyonic Black Holes with String-Loop Corrections }
\vspace{2cm}
\centerline{\bf Mikhail Z. Iofa \footnote{Skobeltsyn Institute of 
Nuclear Physics, Moscow State University, Moscow 119992, Russia, e-mail: 
iofa@theory.sinp.msu.ru}} 
\centerline{Skobeltsyn Institute of Nuclear Physics}
\centerline{Moscow State University}
\centerline{Moscow 119992, Russia}
\centerline{\today}
\vspace{3cm}
\begin{abstract}
In heterotic string theory compactified to four dimensions
with N=2 supersymmetry, string-loop corrections to the
universal sector of the low-energy effective action  are studied. 
Within the framework of N=2 supersymmetric formulation
of the theory,  in the first order in
string coupling constant, we solve the system of the loop-corrected 
Maxwell and Killing spinor equations. Taking as the in-put the tree-level
dyonic black hole solution, we calculate string-loop corrections to 
the string tree-level metric and moduli of dyonic black hole.
\end{abstract}
\vspace{1cm}
\vspace*{\fill}
\bigskip
\normalsize 
\section{Introduction}

At present, string theory is considered the best candidate for a
fundamental theory that would be a consistent quantum theory of
gravity unified with the other interactions~\cite{gsw}.  In
particular, in string theory a large number of classical solutions with 
horizons was found~\cite{dy,t1}
and refs. therein. In this
setting, we meet a fundamental problem of understanding how the
intrinsically stringy quantum effects modify the Einstein gravity and 
classical black-hole
solutions of equations of motion.

In perturbative approach to string theory,  the amplitudes are calculated
as a sum of contributions from the string world sheets
of different genera (for example, \cite{gsw,hp}).
The effective action, except for gravity and matter, contains
 a number of fields inherent to string theory,
such as tensor fields, dilaton and other moduli.  The string 
tree-level effective 
action of the light modes receives both  $\a'$ and
string-loop corrections from higher genera (see, for example, the review
\cite{t2}). 
Thus,  the standard Einstein gravity 
and black-hole solutions  
considered within the framework of superstring theory are also modified by
the string-loop corrections. 

In this paper, we study the  string-loop corrections to the 
tree-level black hole dyonic solution of the equations of motion of the 
$N=2$ supersymmetric $4D$ theory obtained by a suitable compactification
of the heterotic string theory. 
In distinction to the systems with $N=4$ supersymmetry,
where the loop corrections to the two-derivative terms vanish, 
and the models with $N=1$ supersymmetry,
where the amplitudes receive the infinite 
number of loop corrections \cite{agn}, in  models with $N=2$ supersymmetry 
loop corrections to the relevant objects vanish beyond one loop, which makes
 the problem  of their account more or less tractable cf.~\cite{dkl,kou,kir} 
and refs. therein.      

Calculation of the
string-loop corrections  to the two-derivative  terms of
the  bosonic part of the $N=2$ supersymmetric heterotic string
tree-level effective action from the world
sheets of torus topology in the path-integral approach shows
that the Einstein term receives no correction, while the
gauge couplings and the moduli part of the action are modified cf.~\cite{kir}.
Instead of performing rather complicated explicit path integral
calculations to different terms of the action,
following the usual practice,
we make use of the general approach to $N=2$ locally
supersymmetric theories \cite{dewit} and refs. therein. 
Tree-level heterotic string 
effective action can be obtained from the prepotential of the $N=2$
supersymmetric $STU$ model \cite{duff,wikalu,b6}. 
Loop corrections to the tree-level prepotential of the $N=2$
supersymmetric heterotic string effective action vanish
beyond the one loop. Using the loop-corrected prepotential, it is possible
to calculate the loop corrections to various
terms of the effective action in  a universal way.

To be concrete, we consider the effective action  of the
heterotic string theory compactified on the manifold $K3\times T^2$ 
or on a suitable orbifold
yielding   $N=2$ local supersymmetry in $4D$.

To obtain classical solutions  of supersymmetric theory, it is possible either
to solve the second-order equations of motion, or Killing spinor equations
which are conditions that supersymmetry variations
of the fermionic components of the superfieldes to vanish. 
In this paper, to calculate the loop corrections to classical solutions,
 we follow the second way, solving the system
of Killing spinor and Maxwell equations (cf. \cite{mi1} and \cite{mi2}
 where the
loop corrections to magnetic black hole 
were calculated by solving the second-order equations
equations of motion and Killing spinor equations).

 We consider the universal sector of the
theory which contains gravity, the moduli connected with the metric $G_{mn},
\, m,n =1,2 $ and the antisymmetric tensor $B_{12}$
of the two-torus $T^2$, and four vector fields which are expressed through
the mixed internal-space-time components $G_{m\n}$ and $B_{m\n}$ of the
metric and antisymmetric tensor and study  dyonic solution 
of $N=2$ supersymmetric $4D$ theory with two
electric and two magnetic charges.

In the first order in string coupling constant $\e = e^{\p_{\infty}}$, where
$\p_{\infty}$ is the asymptotic value of the dilaton, we find the explicit form
of the loop corrections to the tree-level metric and moduli of a general
dyonic solution.  

 To simplify solution of the loop-corrected equations, we
consider a particular tree-level dyonic black-hole solution with the
constant metric $G_{mn}$ of the
two-torus $T^2$ and vanishing antisymmetric tensor $B_{12}$. 
With this choice, the
standard tree-level moduli $T$ and $U$ are independent of
coordinates, but dilaton and space-time metric are coordinates-dependent.
 This technical simplification makes
possible to avoid too cumbersome expressions for solutions to the system of
the Maxwell and  Killing spinor equations and to obtain
explicit expressions for the loop corrections to the metric and moduli.

Loop corrections to the metric and dilaton are expressed via the
unambiguous real part of the loop correction
to the prepotential $Re\,h (T,U,\bar{T},\bar{U})$. The loop-corrected dyonic metric is
$$g_{00}=-g^{ii} = \frac{r^2}{(r+P)(r+Q)}\left(1 
-\e\frac{PH}{r+Q}\right),$$
where $H=Re\,h/2TU =Re\,h/2G_{11},\, P =(8 P^0 P^1)^{1/2},\, 
Q =(8Q_2 Q_3 )^{1/2}$ 
and $ P^0 , P^1$ and $Q_2 , Q_3 $ are electric and magnetic charges of the
dyonic black hole, $T$ and $U$ are the standard moduli calculated with the
tree-level dyonic solution.   
Loop corrections to the two-torus metric are 
$$ \delta G_{11} =\e C_1\frac{r}{r+P},\quad \delta G_{22}=
\e\left( C_2\frac{r}{r+Q} -(L_2 -L_3 )\frac{P}{r+Q}\right)
$$
and are finite for all $r$. Here $C_i$ are arbitrary
constants, $L_2 =\pa_T Re\,h /U$ and $L_3 =\pa_U Re\,h/T$.

In the case of magnetic black hole, the
string-tree-level dilaton  increases
at small distances from the origin, and the loop corrections
 blow up at the origin. In electric case  and for a
general dyonic
solution  with two electric and two magnetic charges, dilaton is finite
in the whole space-time, and corrections to the tree-level
metric and moduli  are  coordinate-dependent, but finite. 

Although the tree-level dyonic solution is static, with vanishing
axions (imaginary parts of the moduli) and four charges,  in the
loop-corrected solution, generically, can appear new fields
of the first order in string
coupling, in particular, axions and new
vector fields. At the same time, generically, 
the loop-corrected solution becomes
stationary, with the stationarity coefficients of the first order in
string coupling. 
However, for a suitable choice of arbitrary constants
 there exists a stationary loop-corrected solution with four
charges, but with  non-vanishing axions.

In Sect.2 we review direct path-integral calculation
of string-loop corrections to various terms of the universal sector of the
low-energy effective action performing integration over the world sheets
of torus topology.

In Sect.3
we discuss formulation of the theory based on $N=2$ local supersymmetry.
In this  approach, dynamics of the
theory is defined in terms of the prepotential of the theory.
The prepotential of $N=2$ locally supersymmetric theory receives 
string-loop correction only from the world sheets of torus
topology \cite{wikalu,afgnt,hm,defekoz}. 
Using the loop-corrected prepotential, we calculate the gauge 
couplings
in the first order in string coupling and the ambiguities in the gauge
couplings due to the ambiguity in the loop correction to the prepotential. 
In perturbative approach, corrections to the gauge couplings and to the
K\"{a}hler potential for the moduli, which are of the first order in string
coupling, are calculated by substituting the tree-level moduli.

We find the form of the loop-corrected symplectic transformation 
connecting the holomorphic section with the prepotential and that associated
with the heterotic string compactification.

In Sect.4 we  derive the Killing
spinor equations for the gravitino and the fermionic superpartners
of the moduli 
and present them in different forms convenient for subsequent solution.
  
In Sect.5, solving the combined system of the Killing spinor and Maxwell
equations we obtain the tree-level dyonic black hole which
in the following we use as the in-put in calculations of the
loop corrections.

 In Sect.6, in the first order in string coupling constant, we obtain
 the loop-corrected solution for the field strengths, metric and moduli. 
Although the gauge couplings contain the ambiguities,
in the field strengths they cancel.
 
In Sect.7 we consider the loop-corrected
expressions for the BPS and ADM masses
of the black hole. As expected in supersymmetric theories, for solutions
with partially broken supersymmetry the ADM mass obtained
from the asymptotic form of the metric is equal to the BPS mass.
\newline
In Sect.8  we discuss the equations for the axions which are the imaginary
parts of the gaugini Killing spinor equations. The metric components
responsible for non-stationarity of solution are expressed as functions of
the axions. It is shown that there is a solution for the
axions for which the metric remains static.
\section{String-loop corrected $N=2$ supersymmetric effective action}
The bosonic part of the universal sector of the $6D$ effective action of 
the heterotic string theory compactified to six dimensions
with $N=1$ supersymmetry on the 
manifold $K3$ or on a suitable orbifold  is 
\be   
I_6 =\frac{1}{2{\kappa_6}^2}\int d^6x \sqrt{-G^{(6)}}e^{-{\Phi}'}\left
[R^{(6)} + (\partial {\Phi}')^2 - \frac{H^2}{12} \right ]+\ldots.
\label{E1}
\ee
Further compactification on  a two-torus yields the $4D$ theory with $N=2$
 supersymmetry. 
The standard decomposition of the $6D$ metric is 
\be
G^{(6)}=\left(\begin{array}{cc}
 G_{\m\n} + A^m_\m A^n_\n G_{mn}& A^m_\m G_{mn} \\
A^n_\n G_{mn} & G_{mn} 
\end{array}\right).
\ee   

Dimensional reduction of the action (\ref{E1}) yields \cite{sen}
\begin{equation}
\label{E2}
I_4=\frac{1}{2{\kappa_4}^2}\int d^4 x\sqrt{-{G'}^{(4)}}e^{-{\p'}}\left [R(G') +
(\partial {\p'} )^2 - \frac{(H')^2}{12}+{1\over4} \F(LML)\F +\frac{1}{8}Tr
(\partial ML\partial ML) \right ],
\end{equation}
where $\m,\n = 0,\dots,3$ and $m,n =1,2$. The second pair of the vector
fields are the components $B_{m\m}$ of the antisymmetric field $B^{(6)}$.

Here
\be
\label{B2}   
\begin{array}{ll}
M=\left (
\begin{array}{cc}
\,G^{-1} & G^{-1}B\\
-BG^{-1} & \,G 
\end{array}    
\right ),  \qquad
L=\left (
\begin{array}{ll}
0 & I_2 \\
I_2 & \,0 
\end{array} \right )
\end{array},
\ee
and
\ba
G'&\equiv & G'_{\m\n}= G_{\m\n} + A^m_\m A^n_\n G_{mn}, \qquad G\equiv
G_{mn},\nonumber \\
H'&\equiv & H'_{\m\n\l} = H_{\m\n\l} -(A^{(1)n}H_{n\n\l} -A^{(1)m}A^{(1)n}H_{mn\l} +
cycl. perms.).\nonumber
\ea
A direct way to calculate corrections to the string-tree-level effective action 
from integration over the string world
sheets of torus topology is to perform the path integrals for correlators
yielding the relevant structures in the effective action.
For heterotic string theory, The part of the world-sheet
action of the heterotic string theory which
depends on background fields from the universal sector is
\be
\label{D9}
I_{1,0} =\int\, d^2 z d\vt(G_{MN} +B_{MN})(\h X )D{\h X}^M \bar{\pa}{\h X}^N.
\ee
Here $I_{1,0}$ is the action with the $(1,0)$ supersymmetry in the left
supersymmetric sector, ${\h X}^M =X^M +\vt\psi^M$, $M=0,...,9$,
$D=\pa_\vt+\vt\pa_z$.

Performing integration over $\vt$ one arrives at the action
\ba
\label{D9a}
I_{1,0}& =&\int\, d^2 z (G_{\m\n} + B_{\m\n} )(\pa X^\m \bar\pa X^\n
-\k^\m \bar\pa \k^\n ) +
(G_{\m n} + B_{\m n} )(\pa X^\m \bar\pa X^ n
-\k^\m \bar\pa \k^ n )  \\\nonumber
+(G_{m\n}& +& B_{m\n} )(\pa X^m \bar\pa X^\n
-\k^m \bar\pa \k^\n ) + 
(G_{m n} + B_{m n} )(\pa X^m \bar\pa X^n
-\k^m \bar\pa \k^n ) \\\nonumber
+(G_{\m\n}& +& B_{\m\n} )_{,\,\r}\k^\r \k^\m  \bar\pa X^\n +
(G_{\m n} + B_{\m n} )_{,\,\r}\k^\r \k^\m  \bar\pa X^ n +
(G_{m\n} + B_{m\n} )_{,\,\r}\k^\r \k^m  \bar\pa X^\n  \\\nonumber
{}&{}&+(G_{mn,\r }+B_{mn,\r })\k^\r \k^m  \bar\pa X^n .
\ea
To be concrete, we have in view  heterotic string theory compactified to four 
dimensions on the orbifold $T^4 / Z_2 \times T^2 $. The partition function 
has the form of the 
sum of terms, where each term is the product of contributions from 
integration over bosonic and fermionic variables.

Correlator of free bosons on the world sheet of torus topology
with the Teichmuller parameter $\tau$ is given by \cite{hp}
\be
\label{D2}
<X(z.\bar z )X(0)>=-\log|\vt_1 (z)|^2 -\frac{\pi}{2\tau_2}(Im z)^2
\ee
Correlator of fermions with even spin structures $a,b \neq 1,1$ is
\footnote{We consider only fermions with even spin structures 
which yield non-vanishing contributions to the correlators.}
\be
\label{D3}
S{a\atopwithdelims[]b}(z) =<\psi
(z)\psi(0)>=\frac{\vt{a\atopwithdelims[]b}(z){\vt}'_1(0)}
{\vt{a\atopwithdelims[]b}(0)\vt_1 (z)},
\ee
where $\vt{1\atopwithdelims[]1}=\vt_1$.

Contribution from the integration over the left-moving fermions 
 (in the light-cone gauge) yields the blocks of the
form (see, for example, \cite{kir})
\be
\label{D1}
\sum_{a,b=0}^1 (-)^{a+b+ab}\frac{\vt^2 {a\atopwithdelims[]b} (0)
\vt {a+h\atopwithdelims[]b+g} (0)\vt {a-h\atopwithdelims[]b-g} 
(0)}{\eta^4}
\ee
In untwisted case, eight left-moving fermions have the same spin 
structures, and the resulting $4D$ theory has $N=4$ supersymmetry. 
Shifts $h,g$, taking values $0,1$, appear from the orbifold construction
 and reduce supersymmetry to $N=2$.

Performing in (\ref{D1}) summation over spin structures and using the 
Jacobi identities, one has
\be
\label{D4}
{1\over2}\sum_{a,b=0}^1 (-)^{a+b+ab}\frac{\vt^2 {a\atopwithdelims[]b} (0)
\vt {a+h\atopwithdelims[]b+g} (0)\vt {a-h\atopwithdelims[]b-g}
(0)}{\eta^4}=
-\vt^2{1\atopwithdelims[]1}\vt{1-h\atopwithdelims[]1-g}
\vt{1+h\atopwithdelims[]1+g}(0)=0,
\ee
because $\vt_1 (0)=0$.

Let us show that the Einstein term receives no corrections from the world sheets
of torus topology. The graviton vertex function is \cite{gsw,hp,kir}
\be
\label{D5}
V_G =\int \,d^2 z\, G_{\m\n}(X)(\pa X^\m +i(p\psi)\psi^\m )\bar{\pa}X^\n 
e^{ipX}.
\ee
Let us calculate the two-point correlator of the vertex functions 
(\ref{D5}) and take the $O(p^2)$ piece
\footnote{It is understood that  the wave 
function factors making the amplitude vanishing on-shell are removed.}.
Due to (\ref{D4}), purely bosonic part of the correlator $<V_G\,V_G >$ vanishes. The 
four-fermion part is proportional to the expression
\be
\label{D6}
{1\over2}\sum_{a,b\neq 1,1}^1 (-)^{a+b+ab}\frac{\vt^2 
{a\atopwithdelims[]b} (0)
\vt {a+h\atopwithdelims[]b+g} (0)\vt {a-h\atopwithdelims[]b-g}
(0)}{\eta^4}S^2{a\atopwithdelims[]b} (z)=4\pi^2\eta^2 
\vt{1-h\atopwithdelims[]1-g} \vt{1+h\atopwithdelims[]1+g}(0),
\ee
where again we used the Jacobi identity and the relation $\vt'_1 =
-2\pi \eta^3$.
Expression (\ref{D6}) is independent of $z$. The integral over the
 $z$-dependent part of the correlator is
\be
\label{D7}
\int\,d^2 z <\bar{\pa}X(z)\bar{\pa}X(0)> =-\int\,d^2 z \left(\bar{\pa}^2
\log\bar{\vt}_1 +\frac{\pi}{\tau_2}\right) =0.
\ee
Thus, the $O(p^2)$ piece of the correlator $<V_G\,V_G>$ vanishes, and there
is no correction to the Einstein term. The same is true for the vertex $V_B$
with $G_{\m\n}$ substituted by  $B_{\m\n}$.

Let us consider the correlators of the gauge  vertices \cite{hp,dkl,kir}
\be
\label{D8}
V_A= \int\,d^2 z \, G_{pq}(X)\,\left(A_\m^p (X) \pa X^\mu +{1\over2}
F^p_{\m\n}\psi^\m\psi^\n \right)\bar{\pa}X^q e^{ipX}
\ee
(the same with $A^p_\m$ substituted by $B_{p\m}$), where
$p,q =1,2$ label the directions of the untwisted torus $T^2$.
Non-vanishing contribution of the left-moving fields is produced by the
fermionic terms in the vertices $V_A$. In contrast to the graviton vertex,
in the present case, the right-moving bosons $X^p$
carry not $4D$, but an internal $T^2$ index.   
The fields $X^p$ can be split into the classical 
\be
\label{D91}
{\tilde X}^p (m,n)= \pi R\left[(m^p -n^p \bar\tau)\frac{z}{i\tau_2}-
(m^p -n^p \tau)\frac{\bar z}{i\tau_2}\right]
\ee
and a quantum parts $Y^p$: $X^p ={\tilde X}^p +Y^p $. Here it is assumed that
the classical fields ${\tilde X}^p $ take values on the circles of the radius
$R$: $\tilde{X}^p \sim {\tilde X}^p + 2\pi Rk$. Contribution of the
right-moving bosons is
\be
\label{D10}
\sum_{(m,n)} e^{-S(m,n)-I(Y )}\left[\bar\pa {\tilde X}^p \bar\pa {\tilde
X}^q + <\bar\pa Y^p \bar\pa Y^q >\right],
\ee
where the classical action is
$$ S(m,n) =\frac{\pi R^2 G_{pq}}{\tau_2}(m^p -n^p \bar\tau)(m^q -n^q
\tau)      
$$
and $I(Y)= \int d^2 z \,G_{pq}\pa Y^p \bar{\pa}Y^q $. 
As in the case of the graviton vertex, the
integral over bosonic correlator $ <~\bar\pa Y^p \bar\pa Y^q>$ vanishes. 
The classical part
is non-zero and produces a one-loop correction to the tree-level term
$G_{pq}F^p_{\m\n}F^{q\,\m\n}$.

In the following, instead of performing  path-integral calculations
 of the string-loop  corrections to various terms of the action
\footnote{The main difficulty in these calculations is
performing the sum over classical configurations in (\ref{D10}) (cf. with
calculations in \cite{dkl}, where, in distinction to the present case, the
pre-exponent factor was $\pa {\tilde X}^p \bar\pa {\tilde
X}^q $). },  
we shall make use of the $N=2$ formulation of the
theory, in which different loop corrections are expressed via the
loop-corrected prepotential.
\section{N=2 formulation of the effective field theory}
\subsection{Heterotic string effective action and prepotential
of the STU model}
The dilaton $\p'$ in (\ref{E2}) can be split into the sum of the constant part
$\p_\infty$ and a
term vanishing at spatial infinity $\phi' =\phi_\infty +\phi$. In string
perturbation theory, higher order
contributions enter with the factor
$e^{{1\over2}\chi {\phi}}$, where $\chi$ is the Euler characteristic of
the string world sheet \cite{ts}.
The exponent $e^{\phi_\infty}\equiv \e $ can be considered as a
string-loop expansion parameter.

 Written in the Einstein frame, where $g_{\m\n}=
e^{-\p}G'_{\m\n}$, the action (\ref{E2}) is \cite{sen}
\begin{equation}
\label{P1}
I_4 = \int d^4 x \sqrt{-g}\left[ R  
-{1\over2}(\partial {\p} )^2 -{e^{-\p}\over4} \F(LML)\F +
\frac{a_1}{4\sqrt{-g}}\F L\z\F +\frac{1}{8}Tr
(\partial ML\partial ML)+\ldots \right ].
\end{equation}
The $4D$ dilaton $\p$ and axion $a_1$ are defined as   
\be
\label{B10}
\p =\Phi -{1\over2}\ln\det(G_{mn}), \qquad
 \pa_\rho a_1 = -{H'}^{\m\n\l }e^{-2\p}\sqrt{-g}e_{\m\n\l\rho }. 
\ee

The bosonic part of the $4D$ action  written in the standard form of $N=2$ 
supersymmetric theory is 
(for example, \cite{andr,cafpr,4a,mo} and references therein)
\begin{equation}
\label{E4}
I_4^{N=2} = \int d^4 x \sqrt{-g}\left[\frac{1}{2} R + (\bar{N}_{IJ}
\F^{-I}\F^{-J}-
N_{IJ}\F^{+I}\F^{+J} ) + k_{i\bar{j}} \pd y_i \pu \bar{y}_j +\ldots\right], 
\end{equation}
where
$$  \F^{\pm }_{\m \n} =\frac{1}{2}( \F_{\m\n} \pm i\sqrt{-g}\z
\F_{\m\n}) =\frac{1}{2}(\F_{\m\n} \pm\frac{i}{2} \e_{\m\n\r\l}\F^{\r\l}) . $$
Here $\z \F_{\m\n} ={1\over2}e_{\m\n\r\l}\F^{\r\l}$, where $e_{\m\n\r\l}$ is
the flat antisymmetric tensor, $e_{0123} =-1$.
 The gauge coupling constants  $N_{IJ}$ are defined                      
below.
The K\"{a}hler metric is
$$ k_{i\bar{j}}=\frac{\pa^2 K^{(0)}}{\pa y_i \pa \bar{y}_j }, $$
where $y_i$ are the tree-level moduli, and the tree-level 
K\"{a}hler potential  $K^{(0)}$ is
$$ K^{(0)} =-\log(y_1 +\bar{y}_1 )(y_2 +\bar{y}_2 )(y_3 +\bar{y}_3 ).
$$

In the case of the $N=2$ supersymmetric compactification of
superstring theory dynamics of the theory is encoded in the prepotential
which tree-level part is
\begin{equation}
F=-\frac{X^1 X^2 X^3}{X^0} +\ldots ,
\label{E5} 
\end{equation}
where 
\begin{eqnarray}
\frac{X^1 }{X^0 } =iy_1 =i \left(e^{-\p } +ia_1 \right ), \nonumber \\ 
\frac{X^2 }{X^0 } =iy_2 =i \left(e^{\g +\s} +ia_2 \right ),\nonumber \\
\frac{X^3 }{X^0 } =iy_3 =i \left(e^{\g -\s} +ia_3 \right )
\label{E6}
\end{eqnarray}
and dots stand for contributions from other moduli. Here and below $I,J
=0,\ldots, 3$ and $i,j =1,2,3$.

The moduli $y_i$ are equal to conventional moduli $S,T,U$:
\be
\label{E6a}
(y_1 ,y_2 ,y_3 )=(S=e^{-\p } +ia_1, \,T=\sqrt{G}+iB_{12},
\,U=\frac{(\sqrt{G}+iG_{12})}{G_{22}} ).
\ee
The axion $a_1$ is defined in (\ref{B10}),
$a_2 =B_{12}$, and $\g ,\s $ and $a_3$ are read off 
 by comparing definitions (\ref{E6}) with parametrization of 
the metric components of the two-torus
\be
G_{mn}=e^{2\s}\left(\begin{array}{cc}
e^{2\g -2\s} +a_3^2 & -a_3 \\
-a_3 & 1
\end{array}\right).
\label{E3}
\ee

The gauge part of the action (\ref{E2}) with $B_{12}=0$ and diagonal metric 
$G_{mn}$ is
\be
\label{E7}
-\frac{1}{4}G_{11} (\F^{(1)1})^2-\frac{1}{4}G_{22} (\F^{(1)2})^2
 - \frac{1}{4}G^{11} (\F^{(2)}_1)^2 -\frac{1}{4}G^{22} (\F^{(2)}_2)^2,
\ee
where
$$ \F^{(1)m}_{\m\n} = \pa_\m A^m_\n -\pa_n A^m_\m,\quad
 \F^{(2)}_{m\m\n} = \pa_\m B_{m\n} -\pa_n B_{m\m}.$$

It is convenient to relabel the vector fields in correspondence with the
moduli with which they form the superfields
\be
A^1_\m =\sqrt{8}\A^0_\m,  \quad B_{1\m} =\sqrt{8}\A^1_\m , \quad
A^2_\m =\sqrt{8}\A^2_\m \quad B_{2\m}=\sqrt{8}\A^3_\m .
\label{E8}
\ee
The factor $\sqrt{8}$ appears because of different normalizations of the
gauge field in the actions (\ref{P1}) and (\ref{E4}).
\subsection{Loop-corrected gauge couplings }
In the case of the $N=2$ supersymmetric compactification of
superstring theory, the prepotential receives only
only one-string-loop correction 
 \cite{wikalu,b6,afgnt,mo}
\begin{equation}         
F=-\frac{X^1 X^2 X^3}{X^0} -i\e {X^0}^2\e h(-i\frac{X^2}{X^0},
-i\frac{X^3}{X^0})+\ldots,
\label{E5a}               
\end{equation}   
which is independent of the modulus $X^1$.
In the following, we do not rely on the explicit form of the loop correction 
calculated in \cite{kou,afgnt,hm}.

The K\"{a}hler potential becomes
\be
K=-\ln[(y_1 +\bar{y}_1 +\e V)(y_2 +\bar{y}_2)(y_3 +\bar{y}_3)],
\label{E13}
\ee
where the Green-Schwarz function
\cite{kou,wikalu,afgnt,defekoz}
\be
V(y_2,\bar{y}_2,y_3,\bar{y}_3) = \frac{Re\,h -Re\,y_2 Re\,\pa_{y_2 }
 h-Re\, y_3 Re\,\pa_{y_3 } h}{Re\,y_2 \,Re\,y_3}
\label{E14}
\ee
is of the first order in string coupling.
Beyond the tree level, the modulus $S=y_1$ mixes
with other moduli, and its one-loop form is \cite{defekoz}
$$ S = e^{-\p} -\e\frac{V}{2} +ia_1 .
$$

In sections which admit the prepotential, the gauge coupling constants in the
action (\ref{E4}) are calculated  using the formula \cite{dewit}
\be
N_{IJ} = \bar{F}_{IJ} +2i \frac{(Im F_{IK}\, X^K ) (Im F_{JL}\, X^L )}
{(X^I\, Im F_{IJ}\,X^J )},
\label{E9}
\ee
where $F_I =\pa_{X^I}F, F_{IJ}=\pa^2_{X^I X^J}F$, etc. 

Having in view application of general formulas to  dyonic black
hole solutions in which case the tree-level
moduli $y_i$ are real, in the first order in
string coupling constant, we obtain the gauge couplings  $N_{IJ}$ as
\ba
\label{E38} 
&{}&N_{00} =iy^3\left(-1+\e\frac{n}{4y^3 }\right),\quad
N_{01}=-\e\frac{n+2v}{4y_1 }
-i\e a_1\frac{y_2 y_3}{y_1 },\\\nonumber \vspace{2mm}
&{}&N_{02}=-\e\frac{n+2v-2y_2 h y +4y_2 h_2 }{4y_2 }
-i\e a_2\frac{y_1 y_3}{y_2 }, 
\\\nonumber \vspace{2mm}
&{}&N_{03}=-\e\frac{n+2v-2y_3 h y +4y_3 h_3 }{4y_3 }
-i\e a_3\frac{y_1 y_2}{y_3 },\\\nonumber \vspace{2mm}
&{}&N_{11}=-i\frac{y^3}{y_1^2 }\left(1+\e\frac{n}{4y^3 }\right),\quad
N_{12}=i\e y_3 \frac{2y_2 h y-n}{4y^3 } +\e a_3 ,\quad
N_{13} =i\e y_2 \frac{2y_3 h y-n}{4y^3 } +\e a_2 ,
\nonumber \\ \vspace{2mm}
&{}&N_{22}= -i\frac{y^3}{y_2^2 }\left(1-\e\left(\frac{y_2 h_{23}y_3 }{y^3 }+
\frac{n}{4y^3 }\right)\right),\quad 
N_{33}= -i\frac{y^3}{y_3^2 }\left(1-\e\left(\frac{y_2 h_{23}y_3 }{y^3 }+
\frac{n}{4y^3 }\right)\right), 
\nonumber \\\vspace{2mm}
&{}&N_{23}= i\e y_1 \frac{2yhy -4y_2 h_{23}y_3 -n}{4y^3} +\e a_1 .
\nonumber
\ea
Here we kept notations $y_i$ for the real parts
of the moduli, the imaginary parts (axions) $a_i$ are of the first order in
the string coupling. The following notations are  introduced 
\be
\label{B9a}
y^3 = y_1 y_2 y_3,\quad hy =h_a y_a =h_2 y_2
+h_3 y_3, \quad h_a =\pa_{y_a} h,\quad h_{ab}=\pa_{y_a}\pa_{y_b}h
\ee  
and
\be
\label{B9}
 v= h-y_a h_a, \qquad n= h- h_a y_a +y_a h_{ab} y_b, \qquad y_2 hy =y_2
h_{2a}y_b.
\ee
The terms of the first order in string coupling $\e$ are linear functions
of the loop correction to the prepotential $h$ and are calculated by
substituting the tree-level moduli obtained from the input tree-level 
dyonic black hole solution \cite{cvyo,bh2} and also \cite{4a,mo,bls,bt} and
refs. therein.
The imaginary parts $a_i$ of the moduli $y_i$ which are absent
in the classical dyonic  solution can appear  in loop-corrected
solutions of field equations in the next
order in  string coupling constant.

Deriving the above expressions we assumed that the function $h(y_2 ,y_3 )$  
is real. Explicit calculation \cite{hm} (see Sect.6.4) 
shows that the function $h$
contains an unambiguous imaginary polynomial part which we combine
with the imaginary ambiguities discussed below.

The loop correction to the prepotential is defined up to
an arbitrary quadratic polynomial $P(y_2 , y_3 )$ 
in variables $y_2 ,y_3$ and $ y_2 y_3 $ with imaginary coefficients 
\cite{afgnt,hm,wikalu}. Application
of the formula (\ref{E9}) shows that for real $y_2 , y_3$ and real function
$h(y_2 , y_3 )$  
the ambiguities of the couplings $N_{IJ}$
contain the extra factor $i$ with respect to the unambiguous part. 

Introducing
$$ b= 4(P_a y_a -y_a P_{ab}y_b ) , \quad b_2 =4(P_2 -y_a P_{a2}), 
\quad b_3 =4(P_3 -y_a P_{a3}), \quad P_{00}=2P-2y_a P_a +y_a P_{ab}y_b,
$$
where we used the same notations as above in  (\ref{B9a}), 
we obtain the ambiguity of the couplings as
\ba
\delta N_{IJ}=
\left(\begin{array}{cccc}
P_{00}+3b & -\frac{ib}{y_1} & \frac{i(2b_{2} y_{2} -b)}{y_2} & 
\frac{i(2b_3 y_3-b)}{y_3}\\[2mm]
-\frac{ib}{y_1}  & \frac{b}{y_1 ^2} & \frac{(b-2b_2 y_2)}{y_1 y_2} & 
\frac{(b -2b_3 y_3)}{y_1 y_3}\\[2mm]
\frac{i(2b_2 y_2 -b)}{y_2} & \frac{(b-2b_2 y_2)}{y_1 y_2}
 &-P_{22}+\frac{(b-4b_2 y_2)}{y_2^2}&-P_{23} -\frac{b}{y_2 y_3}\\[2mm]
\frac{i(2b_3 y_3-b)}{y_3}&\frac{(b -2b_3 y_3)}{y_1  y_3}
 &-P_{23} -\frac{b}{y_2  y_3}&-P_{33}+\frac{(b-4b_3 y_3)}{y_3^2} 
\end{array}\right).
\ea
Note that the gauge couplings of the vector fields in the universal sector
residing from the moduli supermultiplets
have different structure from the Wilsonian couplings of the gauge fields
in the matter sector originating from the terms $\left[f_{ab}W^a W^b
\right]_F$ in the locally supersymmetric action \cite{defekoz}.  
\subsection{Symplectic transformations}
The field equations and the Bianchi identities for the gauge field strengths
are
\ba
\label{E15}
\pd \left( \sqrt{-g} Im \,G^{-\,\m\n }_I \right) =0 \nonumber \\
\pd \left (\sqrt {-g}Im\, \F^{-J\,\m\n} \right) =0
\ea
where $G^{-\,\m\n }_I=\bar{N}_{IJ}\F^{-J\,\m\n}$.
Eqs.(\ref{E15}) and  are invariant under the symplectic $Sp(8,{\bf{Z}})$
transformations  
\be
O=\left(\begin{array}{cc}
A&B \\
C&D
\end{array}\right)
\label{E10}
\ee
where
\be
 A^T C-C^T A=0, \quad B^T D -D^T B =0, \quad A^T D -C^T B=1. 
\label{L1}
\ee
Under symplectic transformations the couplings $N_{IJ}$ are transformed as
\be
\h{N}=(C+D N)(A+B N)^{-1}.
\label{E12}  
\ee

In sections which do not admit a prepotential (including that which
naturally appears in  $4D$ compactification 
of the heterotic string), the gauge
couplings are calculated by making a symplectic transformation of the
couplings (\ref{E38}) calculated in the section with the prepotential. 
Specifically, at the tree level,
the section connected with   compactification of the heterotic string from
$6D$ to $4D$ is obtained from that with the prepotential
(\ref{E5a}) by  symplectic transformation \cite{wikalu}
\be
A=diag (1,0,1,1),\, B=diag (0,1,0,0),\, C=diag (0,-1,0,0),\,
D=diag (1,0,1,1).
\label{E11}
\ee

At the one-loop level, we look for a symplectic transformation connecting
heterotic section with that with the prepotential in the form
\ba
A=diag (1,0,1,1) +\e(a_{ij}),\, B=diag (0,1,0,0)+\e(b_{ij}), \nonumber \\ 
C=diag (0,-1,0,0)+\e(c_{ij}),\,D=diag (1,0,1,1)+\e(d_{ij}).
\label{L2}
\ea
The matrices $a,\,b,\,c$ and $d$ are constrained by relations (\ref{L1}). 

In the case of  tree-level solutions with  vanishing
axionic parts, the tree-level gauge couplings 
$N_{00}=-iy_1 y_2 y_3, \quad N_{22}=-i\frac{y_1 y_3}{y_2}, \quad 
N_{33}=-i\frac{y_1 y_2}{y_3} $
are proportional to $e^{-\p}$, whereas the expression
 $N_{11}=-i\frac{y_2 y_3}{y_1}$ contains the factor $e^{\p}$. 
In the heterotic section, the tree-level gauge couplings are proportional to
$y_1 =e^{-\p}$.
Loop corrections to the tree-level couplings, as well as non-diagonal
couplings absent at the tree level, appear with the extra factor  $\e e^\p$.

We require that the loop-corrected symplectic transformations (\ref{L2})
produce the same structure of loop corrections to the gauge couplings in the
heterotic section as that which appear in path-integral calculation of
loop corrections, i.e. loop corrections have the extra factor $\e e^\p$ 
as compared to the tree-level expressions.

 Because we perform calculations with accuracy up to the terms of the
first order in string coupling, 
corrections to to the gauge couplings due to the one-loop term in the
prepotential and  to the loop corrections to the tree-level
 symplectic transformations can be treated independently.
From the relations (\ref{L1}) we find that admissible non-zero
entries are $c_{ij}$ with $c_{1i} =c_{i1} =0$ and $d_{11}$. 

Calculating the matrix of the loop-corrected gauge couplings in the
heterotic section and requiring that it is of the form discussed above, we
find 
\be
\h{N}_{IJ}=\left(\begin{array}{cccc} 
N_{00}+ \e c_{00}
-{N_{01}^2 \over N_{11}} & {N_{01}\over N_{11}}
& N_{02}+\e c_{02}-{N_{01} N_{12}\over N_{11}}& 
N_{03}+\e c_{03}-{N_{01}N_{13}\over N_{11}}\\ [2mm]
{N_{10} \over N_{11}} & -\frac{1}{N_{11}} +\e d_{11} &
\frac{N_{12}}{N_{11}} &
\frac{N_{13}}{N_{11}}\\[2mm]
N_{20}+\e c_{20} -{N_{21} N_{10} \over N_{11}} &\frac{N_{21}}{N_{11}}
&N_{22}+\e c_{22} -{N_{21}^2 \over N_{11}} & 
N_{23}+\e c_{23} -{N_{21}N_{13} \over N_{11}}\\[2mm]
N_{30}+\e c_{30}-{N_{31}N_{10}\over N_{11}} & \frac{N_{31}}{N_{11}} &
N_{32}+\e c_{23} -{N_{31}N_{12} \over N_{11}}& N_{33} +\e c_{33} -
{N_{31}^2 \over N_{11}}  
\end{array}\right).
\label{E16}
\ee
where $c_{ij}$ and $d_{11}$ are real constants
which shift the topological term in the action and do not
affect the equations of motion.

Since at the tree level there are only diagonal couplings, the non-diagonal
terms are of next order in $\e$. The terms of the form 
$\frac{N_{1I}N_{1J}}{N_{11}}$  are also of the next
order in the string coupling.  

In the basis associated with the heterotic string compactification,
the part of the matrix $\hn_{IJ}$ which contains the
imaginary parts of the moduli  $a_i$ and ambiguities is
\ba
\label{P2} 
\left(\begin{array}{cccc}
P_{00}+3b & a_1 +\frac{b}{y_2 y_3} & -\frac{ia_{2} y_1 y_3 }{y_2}
-\frac{i(b -2b_2 y_2 )}{y_2} & -\frac{ia_{3} y_1 y_2 }{y_3} 
-\frac{i(b -2b_3 y_3)}{y_3}\\[2mm]

a_1 +\frac{b}{y_2 y_3} & -\frac{b}{(y_2 y_3 )^2}& \frac{ia_3 y_1}
{y_2 y_3} +\frac{i(b -2b_2 y_2)}{y_2^2 y_3} & \frac{ia_2 y_1}{y_2 y_3}
 -\frac{i(b -2b_3 y_3)}{y_3^2 y_2} \\[2mm]

- \frac{ia_{2} y_1 y_3 }{y_2}  -\frac{i(b -2b_2 y_2 )}{y_2} & 
\frac{ia_{3} y_1 }{y_2 y_3}
 +\frac{i(b -2b_2 y_2)}{y_2^2  y_3}&-P_{22}-\frac{(4b_2 -b)}{y_2^2}
&a_1-P_{23} -\frac{b}{y_2 y_3}\\[2mm]

- \frac{ia_3 y_1 y_2}{y_3 }-\frac{i(b -2b_3 y_3 )}{y_3}& 
\frac{ia_2 y_1 }{y_2 y_3 }+\frac{i(b -2b_2 y_2 )}{y_3^2 y_2}&
a_1-P_{23} -\frac{b}{y_2 y_3}& P_{33}-\frac{(4b_3 y_3 -b)}{y_3^2 }
\end{array}\right)
\ea

From the symplectic transformation of the field strengths 
\be
\label{E17}
\left(\begin{array}{c}\h\F^- \\ \h{G}^-\end{array}\right)
=O\left(\begin{array}{c} \F^- \\  G^-\end{array}\right)
\ee
we obtain the relation between the field strengths in the section with the
prepotential and those which appear in the heterotic string effective action
in the form
$$ \h{\F}^{-0} = {\F}^{-0} , \qquad \h{\F}^{-2} = {\F}^{-2 }, \qquad
\h{\F}^{-3} ={\F}^{-3}
$$ 
\ba
\label{L5}
\h{G}^-_0& = &G^-_0 +\e (c_{00}\F^{-0} +c_{02}\F^{-2} +c_{03}\F^{-3}),\,\,  
\h{G}^-_1 = -\F^-_1 +\e d_{11}G^{-1},\nonumber \\
 \h{G}^-_2& = &G^-_2 +\e (c_{20}\F^{-0} +c_{22}\F^{-2} +c_{23}\F^{-3}), \,\, 
\h{G}^-_3 = G^-_3 +\e (c_{30}\F^{-0} +c_{32}\F^{-2} +c_{33}\F^{-3}).
\ea
Substituting  expressions for the couplings (\ref{E16}) and  
  the field strengths (\ref{L5})  in
relation $\h{G}^{- }_I=\bar{\hn}_{IJ}\h{F}^{-J}$
 (any $I=0,1,2,3 $ can be used ), we obtain the relation
\be
\label{E18}
{\F}^{-1 } =-\frac{\bar{N}_{10}}{\bar{N}_{11}}\h{\F}^{-0}
+\frac{1}{\bar{N}_{11}}\h{\F}^{-1}
-\frac{\bar{N}_{12}}{\bar{N}_{11}}\h{\F}^{-2} -
\frac{\bar{N}_{13}}{\bar{N}_{11}}\h{\F}^{-3}
\ee
which does not contain the constants $d_{11}$ and $c_{ij}$.
Here and below the entries with hats refer to the section associated with
the heterotic string compactification.

The K\"{a}hler potential is invariant under symplectic transformations. 
\section{Killing spinor equations}
Requiring  the supersymmetry variations of gravitino and gaugini to vanish,
 we obtain a
system of Killing spinor  equations.
To write the supersymmetry transformations, one introduces symplectic
invariants \cite{cafpr,cast,afr}
\ba
\label{E19}
&{}&S_{\m\n}=X^I \,Im\,N_{IJ}\,\F^{-J}_{\m\n},\\\nonumber
&{}&T^-_{\m\n} \equiv 2ie^{K/2} S_{\m\n}
 =2ie^{K/2}X^I Im\,N_{IJ} \F^{-J}_{\m\n} 
\ea
and
\be
\label{E20}
G^{-i}_{\m\n} =-k^{i\bar{j}}\bar{f}^I_{\bar{j}} Im\,N_{IJ} 
\F^{-J}_{\m\n}.
\ee
Here $k^{i\bar{j}}$ is the inverse K\"{a}hler metric, and $f^I_i =
(\pa_{z_i} +\frac{1}{2}\pa_{z_i} K) e^{K/2}X^I$.
Supersymmetry transformations of the chiral gravitino $\psi_{\a\m }$ and
gaugini $\l^i_\a $ are
\ba
\delta \psi_{\a\m} =D_\m \e_\a - T^-_{\m\n}\g^\n \e_{\a\b}\e^\b,\nonumber \\
\delta \l^i_\a = i\g^\m \pd z^i \e^\a +
G^{-i}_{\m\n}\g^\m\g^\n\e^{\a\b}\e_\b,
\label{E21}
\ea
where
$$
D_\m \e_\a = (\pd -{1\over 4}w^{\h{a} \h{b}}_\m \g_{\h{a}} \g_{\h{b}} + 
{i\over 2} Q_\m )\e_\a .
$$
Here $w_\m^{\h{a} \h{b}}$  and $Q_\m =-\frac{i}{2}(\pa_i
K\pa y_i - \pa_{\bar{i}}K\pa \bar{y}_i )$ are the spin and 
K\"{a}hler connections;
$\h{a},\h{b},...$
are the tangent space indices, $a, b, ... $ are the space-time indices.

The metric of a stationary spherically-symmetric             
configuration is
\be
\label{E23}
ds^2 = -e^{2U} (dt +w_i dx^i )^2 + e^{-2U} dx^i dx_i.
\ee
The only non-vanishing components
of the spin connection $w_0^{\h{a} \h{b}}$ are $w_0^{\h{0} \h{b}} =
{1\over 2}\pa_b e^{2U}$ and  $w_0^{\h{a} \h{b}} =
\frac{1}{2}(\pa_a w_b -\pa_b w_a)$.

 We look for a solution of the system of Killing spinor equations with the
supersymmetry parameter satisfying the relation $\e^\a = 
\g_{\h{0}} \e^{\a\b} \e_\b $ \footnote{More exactly,  following
\cite{fre}, one must extract from the supersymmetry parameter the
coordinate-dependent factor.} . 
The $\m =0$ component of Eq.(\ref{E21}) takes the form
\be
\label{C2}
{1\over 4} w_0^{-\,\h{a} \h{b}} \g_{\h{a}}  \g_{\h{b} } \e_\a + 
T^-_{0n}  e^{\h{b}n }\g_{\h{b}} \e_{\a\b } \e^\b =0.
\ee
Here, using antisymmetry of $w_0^{\h{a} \h{b}}$ in upper indices and 
chirality of the spinor $\e_\a$, we transformed $w_0^{\h{a} \h{b}}$ into
$w_0^{-\,\h{a} \h{b}}$. Using the relations  
\be
\label{C3}
G^-_{\h{m}\h{n}} =i e _{\h{m}\h{n}\h{p}\h{0}}G^{-\h{p}\h{0}}, \qquad
 G^-_{\m\n}\g^\m \g^\n \e_\a = 4G^-_{\h{0}\h{n}} \g^{\h{0}} \g^{\h{n}} \e_\a
\ee
valid for any self-dual tensor and chiral spinor,
 Eq.(\ref{C2}) can be rewritten as
\be
\label{E22} 
( w_0^{-\,\h{0} \h{b}}  - T^-_{0n}  e^{\h{b}n })\g_{\h{b}} \e_{\a\b } \e^\b =0.
\ee
 To have a nontrivial solution for
the supersymmetry parameter, we require that
 $$ w_0^{-\,\h{0}\h{b}}-T^-_{0n} e^{\h{b}n} =0,$$
or, taking the real part of this relation,
\be
\label{E24}
 \frac{1}{4}\pa _n e^{2U} - e^{U} Re\, T^-_{0n}  =0.
\ee
Sufficient condition to have a static metric is $Im T^-_{0n}=0$.

Using the relations (\ref{C3}), conditions 
of  gaugini supersymmetry transformation to vanish are written as
\be
\label{E25}
(i\g^n \pa_n z^i \g^{\h{0}} + 4G^{-i}_{\h{0}\h{n}} \g^{\h{0}} \g^{\h{n}})
\e^{\a\b}\e_\b =0.
\ee
Transforming all the $\g$-matrices to the tangent space (or world) indices, 
we find that there is a nontrivial solution provided
\be
\label{E26}
i\pa_n z^i +4e^{-U}G^{-i}_{0n} =0.
\ee
 Contracting the equation (\ref{E26}) with the function $f^I_i$
and using the relation of special $N=2$ geometry
$$
k^{i\bar{j}}f^I_i \bar{f}^J_{\bar{j}} =-{1\over2}(Im N)^{IJ} 
- e^K \bar{X}^I X^J,
$$
 it is obtained in the form (cf.\cite{fk,bls})
\be
\label{E27}
if^I_i \pa_n z^i + 4e^{-U}\left({1\over 2}\F^{-I}_{0n}+ e^K \bar{X}^I
S_{0n}\right) =0.
\ee
Eq.(\ref{E27}) can be recast to a form which contains the fields
$G^{-}_{I\,0n}$ and $\bar{F}_I $. 
Introducing $g_{Ii} = (\pa_i+\frac{1}{2}\pa_i K) e^{K/2}F_I$,
 contracting Eq.(\ref{E27}) with $\bar{F}_I $
and using the identities 
$$\F^{-I} \bar{F}_I = G^{-}_I \bar{X}^I, \quad  
\bar{F}_I f^I_i = \bar{X}^I g_{Ii}$$  
 which follow from
definitions of $f^I_i$ and $G^-_{I\,\m\n}$, we have
$$i\bar{X}^I g_{Ii}\pa_n  z^i + 4e^{-U}\left({1\over 2}G^-_{I\,0n}\bar{X}^I + 
e^K \bar{F}_I \bar{X}^I S_{0n}\right) =0. $$
Removing the functions $\bar{X}^I$, we obtain the symmetric equation
\be
\label{l1}
i g_{Ii}\pa_n  z^i + 4e^{-U}\left({1\over 2}G^{-}_{I\,0n}+ e^K \bar{F}_I S_{0n}\right) 
=0.
\ee
Using the gravitino equation,  Eqs.(\ref{E27}) and (\ref{l1}) are
presented as (cf. \cite{4a})
\ba
\label{l2}
-2\F^{-I}_{0n} =i\left[e^U \pa_n (e^{K/2}X^I )- (e^{K/2}\bar{X}^I )\pa_n e^U
\right]+2Im\,T^-_{0n} e^{K/2}\bar{X}^I -Im (\pa_i K \pa_n y_i )e^{K/2 +U} X^I ,\\
\label{12a}
-2G^-_{I\,0n} =i\left[e^U \pa_n (e^{K/2}F_I )- (e^{K/2}\bar{F}_I )\pa_n e^U 
\right]+2Im\,T^-_{0n} e^{K/2}\bar{F}_I -Im (\pa_i K \pa_n y_i )e^{K/2 +U} F_I.
\ea
Here we used the equality $ \pa_{z^i} K \pa_n z^i  =\frac{1}{2}\pa_n K
+i\,Im (\pa_{y_i} K \pa_n y_i ) $.  Eqs.(\ref{l2}) and (\ref{12a}) 
are not independent, but
one set can be obtained from the other. One can also take some equations
from the first set, and the remaining equations from the second. In this
paper we shall use the first set of Eqs.(\ref{E27}) or (\ref{l2}). Another
 choice useful for practical calculations is to
take for $I=0,1$ the equations from the first set
(\ref{l2}) and for $I=3,4$ from the second (\ref{12a}).  
\section{Solution of the tree-level Killing spinor equations}
In this section, to fix notations for the following,  
we solve the  tree-level system  
of Killing spinor equations for the moduli (cf. \cite{fk,fre,bt}). 
We look for a static solution in the holomorphic section 
associated with the compactified heterotic string with 
 two magnetic fields $\h\F^0_{\m\n}$ and $\h\F^1_{\m\n}$ and two
electric fields $\h\F^2_{\m\n}$ and $\h\F^3_{\m\n}$. We consider the case of
 purely real tree-level moduli $y_i$ (\ref{E6}), i.e. $a_i =0$. 
The moduli in the heterotic holomorphic section are expressed via the moduli 
in the section with the prepotential as \cite{wikalu} 
\be
\label{ca1}
(\h{X}^I ,\h{F}_I ) =
 (1, y_2 y_3 , iy_2 , iy_3 ;\, -iy_1 y_2 y_3 , -iy_1 , y_1 y_3
-\e h_2 , y_1 y_2 -\e h_3 ).
\ee 
Solving the Maxwell equations and  Bianchi identities 
we obtain the tree-level magnetic
\be
\label{E29}
\h\F^{-0}_{0n} = {i\over 2}\frac{\h{P}^0}{\sqrt{-g'}}\frac{x^n}{r} =
{i\over 2}e^{2U} \h{P}^{0} \frac{x^n}{r^3},\qquad
\h\F^{-1}_{0n} = {i\over 2}\frac{\h{P}^1}{\sqrt{-g'}}\frac{x^n}{r}=
{i\over 2}e^{2U} \h{P}^{1} \frac{x^n}{r^3}
\ee
and electric
\ba
\label{E30}
\h\F^{-2}_{0n}=-\frac{\h{Q}_2 }{2\sqrt{-g'}Im\hat{N}_{22} }\frac{x^n}{r},\qquad
\h\F^{-3}_{0n}=-\frac{\h{Q}_3 }{\sqrt{-g'}Im\hat{N}_{33} }\frac{x^n}{r}
\ea
field strengths, where $\sqrt{-g'} = r^2 e^{-2U}$ 
and the gauge couplings  are
$$\hn_{00}=-iy_1 y_2 y_3 ,
\quad \hn_{11}=-i\frac{ y_1}{y_2 y_3},\quad
\hn_{22}=-i\frac{y_1 y_3}{ y_2},\quad \hn_{33}=-i\frac{y_1 y_2}{y_3}.
$$
In the case of real 
moduli $y_i$, comparing with the action (\ref{E1}), we have
$$Im (\hn_{00},\,\hn_{11},\,\hn_{22},\,\hn_{33} )= -e^{-\p} (G_{11},\, G^{11},
\, G_{22},\, G^{22} )
$$
(we remind that hatted expressions refer to the section associated
with the heterotic string compactification).
The charges $\sqrt{8} \h{P}^I$ and  $\sqrt{8} \h{Q}_I$ are constrained to lie 
on an even self-dual lattice \cite{sen}. 
 The tree-level K\"{a}hler potential is
\be
\label{E31}
K=-\ln\, 8y_1 y_2 y_3. 
 \ee
For the symplectic invariant  $S_{0n}$  (\ref{E19}) we obtain
\ba
S_{0n}&=&(Im \hn_{00}\ \h{\F}^{-0}_{0n} + 
y_2 y_3 Im \hn_{11} \h{\F}^{-1}_{0n}+iy_2
Im \hn_{22} \h{\F}^{-2}_{0n}+iy_3 Im \hn_{33}\h{\F}^{-3}_{0n}
)\nonumber \\
&=&
-i\frac{y_1 y_2 y_3 }{2}\left( \h{P}^0 + \frac{\h{P}^1}{y_2 y_3 }
+\frac{\h{Q}_2 }{y_1 y_3 } +
\frac{\h{Q}_3 }{y_1 y_2 }\right)e^{2U}\frac{x^n}{r^3}.  
\ea
Gravitini  Eq.(\ref{E24}) takes the form 
\be
\label{E32}
{1\over 4}\pa_n e^{2U} -  
\left(\frac{y_1 y_2 y_3 }{8}\right)^{1/2}e^{3U}\left( \h{P}^0 + 
\frac{\h{P}^1}{y_2 y_3 }
+\frac{\h{Q}_2 }{y_1 y_3 } +\frac{\h{Q}_3 }{y_1 y_2 }\right)\frac{x^n}{r^3}=0.
 \ee

The tree-level gaugini equations (\ref{E27})  written in the section
associated with the prepotential are
\ba
\label{E33} 
&{}&I=0:\quad \frac{ie^{K/2}}{2}\pa_n \ln y_1 y_2 y_3
-4e^{-U}\left(\frac{1}{2}
\F^{-0}_{0n} +e^K S_{0n}\right)=0\nonumber \\
&{}&I=1:\quad \frac{y_1 e^{K/2}}{2}\pa_n \ln \frac{y_2 y_3}{y_1} +4e^{-U}
\left(\frac{\h{\F}^{-1}_{0n}}{2\bar{N}_{11}} -iy_1 e^K
S_{0n}\right)=0\nonumber \\
&{}&I=2:\quad\frac{y_2 e^{K/2}}{2}\pa_n \ln \frac{y_1 y_3}{ y_2 }
+4e^{-U}\left(\frac{1}{2}\F^{-2}_{0n}
 -iy_2 e^K S_{0n}\right)=0\nonumber \\
&{}&I=3:\quad\frac{y_3 e^{K/2}}{2}\pa_n \ln \frac{y_1 y_2 }{y_3}
+4e^{-U}\left(\frac{1}{2}\F^{-3}_{0n}
 -iy_3 e^K S_{0n}\right)=0
\ea
The field strengths in the section with the prepotential are defined from
those (\ref{E29}) and (\ref{E30}) by using (\ref{L5}) and (\ref{E18}).  

In the following, we  consider a particular extremal dyonic solution of the
Eqs.(\ref{E32})  with arbitrary constant moduli $y_2$ and
$y_3$ and with the charges subject to relations
\be
\label{E35}
 \h{P}^0 = \frac{\h{P}^1 }{y_2 y_3 }, \qquad \h{Q}_2 y_2 =  \h{Q}_3 y_3.
\ee
Four gaugini equations reduce to one equation. The system of the gravitini
and gaugini equations is
\ba
\label{B7}
\pa_n e^{2U} -(8y_1 y_2 y_3 )^{1/2}e^{3U}\left(\h{P}^0 +\frac{\h{Q}_2}{y_1 y_3
}\right)\frac{x^n }{r^3 } =0, \nonumber \\
\pa_n\ln {y_1}-(8y_1 y_2 y_3 )^{1/2} e^U\left(\h{P}^0 - \frac{\h{Q}_2}{y_1 y_3 
}\right)\frac{x^n }{r^3 } =0 .
\ea
Introducing the charges $P$ and $Q$ as
\be
\label{E37}
  P= \sqrt{8 y_2 y_3}\h{P}^0, \qquad Q= \h{Q}_2 \sqrt{\frac{8y_2 }{y_3 }},
\ee
for the metric and dilaton we obtain
\ba
\label{E36}
e^{-2U} = \frac{(P+r)(Q+r)}{r^2 }, \quad 
y_1^{-1} = e^{\p } =\frac{P+r}{Q+r} \equiv \f0
\ea
which is a particular case of a general dyonic BPS saturated
solution \cite{cvyo,bh2}
The components of the metric $G_{mn}$ are
$$G_{11} = y_2 y_3, \qquad G_{22}= \frac{y_2}{y_3}    $$

The factor $\sqrt{8}$ appears because of different 
normalizations of the gauge terms in the actions (\ref{E2}) 
and (\ref{E4}).

 The
charges $\h{P}^{0,1}$ and $\h{Q}_{2,3}$ are expressed
via the charges $P$ and $Q$ as
\be
\label{E47}
\h{P}^0 = \frac{Pe^{-\g_0}}{\sqrt{8}},\qquad \h{P}^1 =
\frac{Pe^{\g_0}}{\sqrt{8}},\qquad
\h{Q}_2 = \frac{Qe^{-\s_0}}{\sqrt{8}},\qquad
\h{Q}_3 = \frac{Qe^{\s_0}}{\sqrt{8}}.
\ee
Here $\g_0$ and $\s_0$ are the tree-level values of the moduli $\g$ and
$\s$ parameterizing the metric of the two torus (\ref{E3}).
\section{Solution of the system of the
loop-corrected Maxwell and Killing spinor equations}
Our next aim is to solve the system of Maxwell and 
Killing spinor equations
 for the loop-corrected metric and moduli using as the input  the 
tree-level solution (\ref{E36}). We look for a solution in the
first order in the string coupling constant. 
The loop corrections to the gauge coupling constants are calculated by
substituting the 
tree-level moduli. The terms which 
depend on the constant tree-level moduli $y_2$ and $y_3$ 
are  independent of coordinates.
Dependence on coordinates enters through the modulus $y_1 ={\f0}^{-1}$ and
the metric.

Let us introduce notations for the loop-corrected metric and moduli.
The functions
$\p, \g $ and $\s$ which enter the moduli (\ref{E3}) are split into the
tree-level
 $\p_0 , \g_0 $ and $\s_0$ and  first-order parts in string
coupling $\p_1 , \g_1 $ and $\s_1$: $\p =\p_0 +\e \p_1$, etc.
 The function $U$ which enters the metric
will be written as $2U_0 +\e u_1 $.
The tree-level expressions for the functions $U_0$ and $\p_0$ are given by
(\ref{E36}), where $\g_0$ and $\s_0$ are arbitrary constants.

\subsection{Electric field strengths}
First let us solve the Maxwell equations (in the heterotic holomorphic section) 
which can be rewritten in the form
\be
\label{B3}  
\pd (\sqrt{-g}\, Im \hn_{IJ} \h\F^J +Re \hn_{IJ}\z \h\F^J )^{\m\n} =0.
\ee
With the  accuracy of the terms of the first order in string coupling $\e$, 
Eqs.(\ref{B3}) written in spherical coordinates are 
\ba
I=0: \pa_r [\sqrt{-g} ( Im \hn_{00} \h\F^0 +Im \hn_{02} \h\F^2 +
Im \hn_{03} \h\F^3 ) + Re \hn_{00}\z\h\F^0 +
Re \hn_{01}\z\h\F^1 ]^{0r} =0 \label{E39}\\
I=1: \pa_r [\sqrt{-g} ( Im \hn_{11} \h\F^1 +Im \hn_{12} \h\F^2 + 
Im \hn_{13} \h\F^3 ) + Re \hn_{10}\z\h\F^0 + Re \hn_{11}\z\h\F^1 ]^{0r} =0 
\label{E40}
\ea
\ba
I=2:\quad \pa_r [\sqrt{-g}(Im \hn_{22} \h\F^2 +Im \hn_{23} \h\F^3 ) + 
Re \hn_{20}\z\h\F^0 + Re \hn_{21}\z\h\F^1  ]^{0r} =0 \label{E41}\\
I=3:\quad \pa_r [\sqrt{-g}(Im \hn_{33} \h\F^3  +Im \hn_{32} \h\F^2 )+ 
Re \hn_{30}\z\h\F^0 + Re \hn_{31}\z\h\F^1  ]^{0r} =0 \label{E42}
\ea
Only the diagonal gauge couplings $\hn_{II}$ contain terms of zero order  in
string coupling. The field strengths $\h\F^{0,1\,0r}$, 
absent at the tree level, are of the first order in the string coupling.  
Substituting the tree-level field strengths (\ref{E29}) and (\ref{E30}) in
Eqs.(\ref{E39}) and (\ref{E40}), we have
\ba 
\label{B5}
\h\F^{0\,0r}& =&\frac{\e\h{q}_0 - Re \hn_{00}\,\h{P}^0-Re\hn_{01}\,\h{P}^1
-\h{Q}_2 \frac{Im \hn_{02}}{Im\hn_{22}}-\h{Q}_3\frac{Im \hn_{03}}{Im\hn_{33}} }
{\sqrt{-g'}\,Im \hn_{00}}, \nonumber \\
\h\F^{1\,0r}& =&\frac{\e\h{q}_1 -Re \hn_{10}\,\h{P}^0 -Re\hn_{11}\, \h{P}^1
-\h{Q}_2\frac{Im \hn_{12}}{Im\hn_{22}}-\h{Q}_3\frac{Im \hn_{13}}{Im\hn_{33}} } 
 {\sqrt{-g'}\,Im \hn_{11}},   
\ea
where $\h{q}_{0,1}$ are arbitrary constants  which have a meaning of 
electric charges  of the first order in string coupling.
 Since the numerators are of the first order in string
coupling, denominators are taken in the leading order. 

In the case of constant moduli $y_2$ and  $y_3$ the  ambiguities
of the couplings in the matrix (\ref{P2}) which are real
 enter the topological terms
in the action and do not affect the equations of motion. The imaginary
ambiguities $\delta \hn_{0a}$ and  $\delta \hn_{1a}$ of the gauge couplings 
appear in the Maxwell equations.
Substituting explicit expressions for the couplings, we obtain
\ba
\label{B13}
\h\F^{0\,0r}&=&\e\frac{\h{q}_0  -a_1\,\h{P}^1 -\h{Q}_2
\left(a_{2} +{(b -2b_2 y_2 )}/{y_1 y_3} \right) 
-\h{Q}_3 \left(a_{3} +{(b -2b_3 y_3)}/{y_1 y_2} \right)} 
{\sqrt{-g'}\,Im \hn_{00}}, \nonumber \\
\h\F^{1\,0r} &=&\e\frac{\h{q}_1 -a_1\,\h{P}^0+ \h{Q}_2 \left(a_3
/y_3^2 -(2b_2 y_2 -b)/{y^3 y_3}\right)+ \h{Q}_3 \left(a_2
/y_2^2 -(2b_3 y_3 -b)/{y^3 y_2}\right) } {\sqrt{-g'}\,Im \hn_{11}}
\ea
For the case we consider with
 the charges subjected to conditions (\ref{E47}) the ambiguities
in the strengths (\ref{B13}) cancel, and the remaining expressions are
unambiguous.    

The Eqs.(\ref{E41}) and (\ref{E42}) yield 
\ba 
\label{B6}
\h\F^{2\,0r}&=&\frac{\h{Q}_2 -Re \hn_{20}\,\h{P}^0
-Re\hn_{21}\,\h{P}^1 -\frac{Im \hn_{23}}{Im \hn_{33}}
\,\h{Q}_3 }{\sqrt{-g'}Im \hn_{22} }\nonumber \\                             
\h\F^{3\,0r}&=&\frac{\h{Q}_3 - Re\hn_{30}\, \h{P}^0                                   
-Re\hn_{31}\, \h{P}^1 -\frac{Im \hn_{32}}{Im \hn_{22}}
\,\h{Q}_2 }{\sqrt{-g'}Im \hn_{33} }.
\ea

At this point there are two options. Either we can allow for appearance of
the  fields absent at the tree level with the charges of the 
first order in string coupling, or we can require that  the loop-corrected
solution, as  the tree level one, contains two electric and two magnetic
fields. Since magnetic fields $\z\h{\F}^{2,3}$ with the charges
$\h{p}_{2,3}$ of the first order in string coupling enter the Maxwell
equations multiplied by the non-diagonal gauge couplings
$N_{IJ}$ which are of the first order in
string coupling and thus cannot be fixed with the accuracy
of the first order in string coupling, we set the charges $p_{2,3}$ equal to
zero, but retain the electric charges $\h{q}_0$ and $\h{q}_1$.  

In the section associated with the prepotential, 
the field strengths can be obtained
either by direct solution of the system of Maxwell equations and Bianchi
identities or by using relations (\ref{L5}) and (\ref{E18}). 

Using  solutions for the field strengths (\ref{B5}) and (\ref{B6}), 
in the first order in the string coupling constant, we
calculate  the symplectic invariant $S_{0n}$ (\ref{E19}) as
\ba
\label{E44}
&{}& S_{0n}=\{[\h{P}^0 (Im\, N_{00} +y_i Re N_{i0}  ) 
 -\h{P}^1 y_1 - ( \h{Q}_2 y_2 + \h{Q}_3 y_3 )]\\\nonumber
&{}&-i\e [\h{P}^0 (a_1 y_2 y_3 +a_2 y_1 y_3 + a_3 y_1 y_2 )
 + \h{P}^1 a_1 + a_a \h{Q}_a  -\h{q}_0  - \h{q}_1  y_2 y_3 )]\}
\frac{i}{2}e^{2U}\frac{x^n }{r^3 }
\ea
Only the the couplings $N_{00}$ and 
$N_{0i}$ , $i=1,2,3$ enter the expression (\ref{E44}). 
Substituting  the loop-corrected gauge couplings (\ref{E38}), we obtain
\be
\label{E45}
Im N_{00} +y_i Re N_{i0} = - y^3 -\e (2v + h_a y_a ) .  
\ee
All the terms containing second derivatives of the loop correction to the
prepotential have canceled.  

Substituting in (\ref{E44}) the expression (\ref{E45}), we finally have
\ba
\label{S}
&{}&S_{0n}=\left\{ [-\h{P}^0 (y^3 +\e (2v + h_a y_a ))  -\h{P}^1
 y_1 - \h{Q}_a y_a]\right. \nonumber \\
&{}&\left.-i\e [\h{P}^0 (a_1 y_2 y_3 +a_2 y_1 y_3 + a_3 y_1 y_2 )
 + \h{P}^1 a_1 + \h{Q}_a a_a
 -\h{q}_0 - \h{q}_1  y_2 y_3  )]\right\} \frac{i}{2}e^{2U}\frac{x^n }{r^3 }
\ea

Because  the Killing spinor
equations (\ref{E26}) are linear in derivatives of the moduli $\pa z^i$,
the equations for the real  parts of the moduli decouple from imaginary parts.
Since Killing spinor equations from which are determined the
loop corrections to the dilaton $\p$ and metric component $e^{2U}$ contain
only the real parts of the moduli, to determine these corrections we can
ignore the imaginary parts of the moduli.
The imaginary parts of the
moduli are of the first order in string coupling, and are determined from
the imaginary parts of the  Killing spinor equations which contain the
real parts only in the main order   
\footnote{To have a static  metric, we must choose a solution for which 
 $Im T^-_{0n}=0$ (see. Sect.3).}. 
\subsection{Loop-corrected gravitini Killing spinor equation}
Expanding  the K\"{a}hler potential (\ref{E13})
to the first order in string coupling, we obtain
\be
\label{E48}
e^K = \frac{\f0 e^{-2\g_0 }}{8}\left[1 +\e\left(\p_1 -2\g_1 \right)\right].  
\ee
Using this expression for the
K\"{a}hler potential and calculating to the first order in
string coupling the real part of the symplectic invariant 
$T^-_{0n} = 2i e^{K/2} S_{0n}$ 
which enters the gravitini equation (\ref{E24})  we have
\be   
\label{E51}
T^-_{0n} = \frac{\f0^{-1/2}}{4}  \left[P\left(1-\e\frac{\p_1 }{2}
+\e H\f0 \right) +Q\f0 \left(1+\e\frac{\p_1}{2}\right) \right]
e^{2U}\frac{x^n }{r^3 } 
\ee
Here we introduced 
\be
\label{E50}
 H= \frac{h}{2}e^{-2\g_0 }. 
\ee
 All the factors  multiplying the expressions of the first order
in string coupling are taken in the main order. 
It is convenient to introduce the functions $q'$ and $l'$ as
\be   
\label{E52}
q' =\frac{{\f0}'}{\f0}=\frac{Q-P}{(r+P)(r+Q)},\quad l'=2U'_0=\frac{2PQ
+Pr+Qr}{r(r+P)(r+Q)}.
\ee
Now the gravitino equation (\ref{E24}) takes the form
\be   
\label{E53}
\frac{u'_1}{l' }-\frac{u_1 }{2} +\frac{\p_1 }{2}\frac{P-Q\f0 }{P+Q\f0 }
-\frac{PH{\f0} }{P+Q\f0 } =0.
\ee
The leading-order terms have canceled due to the Eqs.(\ref{E32}).
\subsection{Loop-corrected gaugini Killing spinor equation}
Let us solve the gaugini  Killing spinor equations in the form (\ref{E27}).
In the first order in string coupling,  the combinations
$f^I_i \pa_n z^i$ 
which enter the equations (\ref{E26}) are
\ba
\label{E46}
 f^0_i \pa_n z^i&=& -{1\over 2}e^{K/2 } \left(1-\e{V\over 2y_1 }\right)\pa_n\ln y_1 y_2
y_3 \nonumber \\
f^I_i \pa_n z^i &=&iy_i \left(B_n^0 +e^{K/2}\pa_n\ln y_i \right). \qquad i=1,2,3,
\ea
With the accuracy up to the terms of order $O(\e )$, the expression
 $f^0_i \pa_n z^i$ is
\be
\label{E55}
f^0_i \pa_n z^i =\frac{ q' {\f0}^{1/2 }e^{-\g_0 } }{2\sqrt{8}}  
\left[1+\e\left(\frac{{\p_1 }'-2{\g_1 }'}{q'}+
\frac{\p_1 -2\g_1 }{2} \right)\right]\frac{x^n}{r}.   
\ee 
Substituting the explicit expressions for the gauge couplings, 
we obtain the combination
$\frac{1}{2}\F^{-0}_{0n} +e^K S_{0n}$ in Eq.(\ref{E27}) as 
\ba   
\label{E54}
\frac{1}{2}\F^{-0}_{0n} +e^K S_{0n}
 = \frac{e^{-\g_0 }}{4\sqrt{8}}\left[
P\left(1+\e\left(\g_1 -H\f0\right)\right) 
 -Q\f0 \left(1+\e\left(\p_1 -\g_1 \right)\right)\right]
{i\over2}e^{2U}\frac{x^n}{r^3}.
\ea
Using Eqs.(\ref{E54}) and (\ref{E55}), we obtain the $I=0$ 
gaugini equation  in the form
\be   
\label{E56}
\frac{\p'_1 -2\g'_1 }{q'} +\frac{\p_1 }{2}\frac{P+Q\f0 }{P-Q\f0 }-\frac{u_1
}{2}-2\g_1 \frac{P}{P-Q\f0 }  +
\frac{PH\f0}{P-Q\f0 } =0 
\ee

Following the same steps, we obtain the $I=1$ gaugini equation as
\be   
\label{E59}
\frac{\p'_1 +2\g'_1 }{q'} +\frac{\p_1 }{2}\frac{P+Q\f0 }{P-Q\f0 }-\frac{u_1
}{2}+2\g_1 \frac{P}{P-Q\f0 } +
\frac{PH\f0 }{P-Q\f0 } =0. 
\ee
 Eqs.(\ref{E56}) and (\ref{E59}) split into the pair of equations
\be   
\label{E60}
\frac{\p'_1 }{q'} +\frac{\p_1 }{2}\frac{P+Q\f0 }{P-Q\f0 }-\frac{u_1}{2}
 +\frac{PH{\f0} }{P-Q\f0 } =0 
\ee
and
\be   
\label{E61}
\frac{\g'_1 }{q'}+\g_1\frac{P}{P-Q\f0 } =0.
\ee
Using the expressions (\ref{E52}) for $q'$ and (\ref{E36}) for $\f0$, we
rewrite  Eq.(\ref{E61}) as
\be   
\label{E611}
{\g_1}' -\g_1 \frac{P}{r(r+P)} =0.
\ee  

Combining  Eqs.(\ref{E27}) with $I=0$ and $I=2$ (the same with $I=3$),
and using the expressions (\ref{E46}) for $B_n^I$, we have
\be   
\label{E64}
ie^{K/2}\frac{\pa_n y_2 }{y_2 } +4e^{-U}\left(\frac{\F^{-2}_{0n}}{2iy_2 }
-\frac{1}{2}\F^{-0}_{0n} -2e^K S_{0n}\right) =0.
\ee 
Since the tree-level moduli $y_2$ and $y_3$ are constants,  $\pa_n y_2$
and  $\pa_n y_3$ are of the first order in string coupling.
The combination $\left(\frac{\F^{-2}_{0n}}{2iy_2 }
-\frac{1}{2}\F^{-0}_{0n} -2e^K S_{0n}\right)$ is also of the first order in
string coupling. Substituting in (\ref{E64}) (the same with $I=3$) 
the explicit expressions, we arrive at
\ba
\label{E65}
{\g_1}' +{\s_1}' +\frac{(\f0)^{-1/2}e^{U_0}}{r^2}\left[P\left((L-L_2)\f0
-\g_1\right)-Q\f0 \s_1 \right] =0,\nonumber \\
{\g_1}' -{\s_1}' +\frac{(\f0)^{-1/2}e^{U_0}}{r^2}\left[P\left((L-L_3)\f0
-\g_1\right)+ Q\f0 \s_1 \right]=0.
\ea
Here we introduced
\be
\label{E63}
L_2 =\frac{h_2}{y_3}, \quad L_3 =\frac{h_3 }{ y_2 },
\quad L=\frac{L_2  +L_3}{2} .
\ee
\subsection{Loop corrections to the moduli and metric}
Combining   Eqs.(\ref{E65}) we obtain
\ba   
\label{E66}
&{}&  {\g_1}' -\g_1 \frac{P}{r(r+P)} =0, \nonumber \\
&{}&{\s_1}' -\frac{\f0^{1/2}e^{U_0} }{r^2}[\frac{1}{2}P(L_3
-L_2 )+Q\s_1 ] =0.
\ea
The first Eq.(\ref{E66}) is identical to Eq.(\ref{E611}). Thus, although the
number of equations exceeds the number of variables, the system of 
equations is consistent.
Substituting explicit expressions (\ref{E36}) for  $\f0$ and $e^{U_0}$, 
we obtain the equation for $\s_1$ in the form
\be
\label{B11}
{\s_1}' -\frac{1}{r(Q+r)}\left[\frac{P}{2}(L_2 -L_3 ) +Q\s_1 \right]=0.
\ee
Solving equations for the loop corrections to the moduli, we obtain
\ba
\label{B12}
&{}&\g_1 = C_1\frac{r}{r+P}, \nonumber\\
&{}&\s_1 =  C_2\frac{r}{r+Q} -\frac{(L_2 -L_3 )}{2}\frac{P}{r+Q},
\ea
where $C_1$ and $C_2$ are arbitrary constants.

Let us solve the system of equations (\ref{E53}) and (\ref{E60}) for the
loop corrections to the metric and dilaton. Using the relation 
\be
\label{E68}
l' =-\frac{P+Q\f0}{P-Q\f0}q',
\ee
we transform the system to the form 
\ba
\label{E69}
&{}&{\p_1}' -\frac{u_1}{2}q' -\frac{\p_1}{2}l' +
\frac{PH{\f0}'}{P-Q\f0}  =0,\nonumber \\
&{}&{u_1}' -\frac{u_1}{2}l' -\frac{\p_1}{2}q'+
\frac{PH{\f0}'}{P-Q\f0} =0.
\ea
We solve the system with the boundary condition that $u_1$ and $\p_1$
vanish in the limit $r=\infty$, i.e. the loop-corrected metric and dilaton
are asymptotic to the Lorentzian metric and constant dilaton equal to unity.
Adding and subtracting  Eqs.(\ref{E69}), we obtain two
separate equations for $u_1 \pm\p_1$ which are straightforwardly solved
as
\be
\label{E70}
u_1 =\p_1 = -\frac{PH}{r+Q},
\ee 
where $H =Re\,h(T,U) /2TU$. Here $h$ is the loop correction to the 
prepotential, $ y_2 =T,\,  y_3 = U$ are the standard (real) moduli (\ref{E6a})
 calculated by substituting the tree-level dyonic solution.

 To be concrete, let us consider 
the explicit form of the prepotential which was calculated in \cite{hm} for the
case of unbroken gauge group is $[E_8\times E_7\times U(1)^2 ]_L
\times U(1)^2_R $. The loop correction is of the form
\ba
h(T,U)&{}& =  Re\,h -i\frac{U^3}{12\pi}+\delta h_+ ,\quad T_2 > U_2 , \\\nonumber
 &{}&   = Re\,h -i\frac{T^3}{12\pi}+\delta h_- ,\quad U_2 > T_2 , 
\label{E71}
\ea
where $Re \,h$ is negative and $\delta h_{\pm}$ are 
ambiguities of the prepotential which are quadratic 
polynomials in $T, U. TU$ with
 purely imaginary coefficients discussed above. 

Since $Re \,h <0$, the function $H$ is negative and the loop corrections 
(\ref{E70}) to the metric and dilaton are positive.  

From the expression for the loop-corrected metric
\be
\label{M1}
e^{2U} = \frac{r^2}{(r+P)(r+Q)}\left(1-\e\frac{PH}{r+Q}\right)
\ee
follows that its domain of validity is 
$1>\e\frac{PH}{r+Q}$. For $Q>|\e PH|$ (\ref{M1}) is valid for all $r$.

In magnetic case, for the purely magnetic solution of the equations of 
motion with loop corrections   the metric is
\be
\label{M2}
e^{2U} = \frac{r}{r+P}\left(1 -\e\frac{PH}{r}\right)
\ee
with the range of validity $r>\e |HP|$ 
\footnote{Generically, loop-corrected solution for magnetic black hole
contains also electric charges of the first order in string coupling.
In this case in (\ref{M2}) appears the expression $\e (PH +Q)$, where $Q$
is a combination of electric charges. This solution cannot be obtained as
the limit $Q\rightarrow 0$ of the dyonic solution, since the limits $r=0$
and $Q=0$ of the function
$\f0$ and other expressions  are not
interchangeable. Since these functions appear in the differential
equations, one must start from the beginning from the purely magnetic
solution. }. At the same time, the in-put
tree-level solution is meaningful in the region where the effective string
coupling constant $\e e^\p$ is small,  $\e \left(1+\frac{P}{r}\right)<1$, or
$r>\e P$. If we extrapolate both the tree-level and the
loop-corrected solutions in the broader region, up to the origin, we see
that the loop correction becomes dominant in this region. The metric
(\ref{M2}) can be considered as the first term
in the expansion of the function $\frac{r+|\e PH|}{r+P}$ which can be
interpreted as a component of the metric with the smeared singularity
at the origin.
\section{BPS mass and asymptotics of the loop-corrected solution}
BPS condition relating the mass of a solution with the central charge
of the $N=2$ supersymmetry algebra should retain its form in perturbation
theory provided the perturbation theory does not violate supersymmetry. The
BPS spectrum of $N=2$ supersymmetric theory is given by \cite{cafpr}
\be
\label{E77} 
M_{BPS} =|Z_{\infty}| = e^{K/2} |n_I X^I - m^I F_I |_{\infty},
\ee
 $n_I$ and $m^I$ are integers
proportional to electric and magnetic charges of the fields 
(\ref{E8}) gauging the group $U(1)^4$. Subscript $\infty$ indicates that the
equality is valid at spatial infinity.

Equality of the ADM and BPS masses of a BPS-saturated solution can be
seen from the Nester construction in which both
masses are expressed via the asymptotics of the function (cf. \cite{duff,gh})
\be
\label{E79}
 T^-_{\m\n} = e^{K/2}(F_I \F^I_{\m\n} -X^I G_{I\m\n} ),
\ee
where $G_{I\m\n} = Re\, N_{IJ}\F^J_{\m\n} -Im\, N_{IJ}\z\F^J_{\m\n} $.
Expression (\ref{E77}) is the integrated asymptotic form of (\ref{E79}).

At the the tree-level, the BPS mass  of dyon with two electric and two
magnetic $m^2 ,m^3$ and two electric charges 
$n_0 , n_1$  calculated  in the heterotic holomorphic section (\ref{ca1}) 
by using the formula (\ref{E77}) reads 
\be
\label{E80}
M_{BPS} =\frac{1}{\sqrt{8e^{2\g_0}}}|n_0 +n_1 e^{2\g_0} +m^2 e^{\g_0 -\s_0} +
m^3 e^{\g_0 +\s_0}|.
\ee
Setting  $(n_0 , n_1 , m^2 , m^3 )=
(\h{P}^1, \h{P}^0 ,\h{Q}^3 , \h{Q}^2)$, we obtain (\ref{E80}) as
\be
\label{E801}
M_{BPS} =\frac{1}{\sqrt{8}}|\h{P}^0 e^{\g_0} +\h{P}^1 e^{-\g_0}  
+\h{Q}_3 e^{-\s_0} +\h{Q}_2 e^{\s_0} |=\frac{1}{4}(P+Q),
\ee
where for the second equality we used (\ref{E47}).
Here we used the fact that the dilaton is normalized at the infinity
 to unity: $y_1|_{\infty} =1$.

On the other hand, calculating the ADM mass from the asymptotics of the
metric, $e^{2U} = 1-\frac{(P+Q)}{r} +O(r^{-2})$, we have $M_{ADM}=2(P+Q)$.
Up to a normalization factor $8$ due to different respective normalization
of the Einstein and gauge terms in the actions (\ref{P1}) and (\ref{E4}),
the ADM and the BPS masses coincide.

At the one-loop level, the BPS mass can be obtained either by using
(\ref{E79}),
with  the loop-corrected (exact, because in $N=2$
supersymmetric theory prepotential receives only one-loop correction)
period vector $(\h{X}^I ,\h{F}_I )$, or from  from the asymptotics of the
loop-corrected  symplectic invariant $T^-_{0n}$ (\ref{E51}) which
enters the Killing spinor equation (\ref{E24}).
We have
\be
\label{ca2}
M_{BPS} = e^{K/2} \left[\h{P}^0 \left(y^3 +\e (2v +h_a  y_a )\right)
+ \h{P}^1 y_1 +
\h{Q}_a y_a \right]_{\infty} + O(\e^2 ),
\ee
or, 
\be
\label{ca21}
M_{BPS} =\frac{1}{4}\left[P(1+\e H )+Q\right].
\ee
The ADM mass is obtained from the asymptotic form of the metric
\be
\label{E82}
e^{2U} |_\infty =e^{2U_0}(1+\e u_1 )|_\infty .
\ee
In the limit $r\rightarrow\infty$ we obtain the asymptotics of the
correction to the metric (\ref{E70})
\be
\label{B8}
u_1|_{r\rightarrow \infty}=-\frac{PH}{r}
\ee
Taking into account the normalization factor $8$, we find that the
loop-corrected 
BPS and ADM masses are equal to each other.  
\section{Equations for axions}
Let us discuss of the
equations for the imaginary parts $a_i$ of the moduli $y_i $.

 Convoluting the
Eq.(\ref{E26}) with the metric $k_{i\bar{j}}$ (summation over $i$), we have
\be
\label{E90}
k_{\bar{j}i}\pa_n (Re\, y_i +i\e a_i ) +
4e^{-U} \bar{f}^I_{\bar{j}} Im\, N_{IJ} \F^{-J}_{0n}=0.
\ee
Introducing $T_{I0r}\equiv Im\, N_{IJ} \F^{-J}_{0r}$ and separating
the imaginary part of Eq.(\ref{E90}), we obtain
\be
\label{E91}
k_{\bar{j}i} a'_i +4e^{-U} Im\, (\bar{f}^I_{\bar{j}} \,T_I )=0
\ee
Since $a_i$ are of the first order in string coupling,  $k_{\bar{j}i}$
should be taken in the leading order.
With the required accuracy, using the expressions in the section
with the prepotential, we calculate
\ba
\label{E92}
T_{0\,0r} &=&\left[-q_0 +iy^3 \h{P}^0 Im N_{00}\right]
\frac{1}{2\sqrt{-g'}}, \nonumber \\
T_{1\,0r}& =&\left[\h{P}^0 Re N_{10}-\h{P}^1 + i(\h{P}^0 Im N_{10}-q_1
Im N_{11} )\right]\frac{1}{2\sqrt{-g'}}, \nonumber \\
T_{2\,0r}& = &\left[Re N_{20} \h{P}^0 -\h{Q}_2 +i Im N_{20} \h{P}^0 \right] 
\frac{1}{2\sqrt{-g'}}, \nonumber \\  
T_{3\,0r}& = &\left[Re N_{30} \h{P}^0 -\h{Q}_3 +i Im N_{30} \h{P}^0 \right]
\frac{1}{2\sqrt{-g'}}.
\ea
In the combination $Im (\bar{f}^I_{\bar{j}} \,T_I )= -Im f^I_j Re T_I +
Re f^I_j Im T_I$ which
enters  Eq.(\ref{E91}) the imaginary parts of the functions $T_I$ with
$I=1,2,3$ and  $Re\, T_0$ are of the first order in string coupling. 
Thus, we need the
corresponding functions $Re\, f^I_j$ and $f^0_j$ in the leading order in
$\e$. 

The functions $Re\,f^I_j$ calculated in the leading order in $\e$ are
\ba
\label{E93}
Re\,f^0_i& =&i\frac{e^{K/2}}{2}\left(\frac{1}{y_1},\frac{1}{y_2},
\frac{1}{y_3} \right),\quad
Re\,f^1_i =\frac{e^{K/2} y_1}{2}\left(\frac{1}{y_1},-\frac{1}{y_2},
-\frac{1}{y_3} \right),
\nonumber \\
Re\,f^2_i& =&\frac{e^{K/2} y_2}{2}\left(-\frac{1}{y_1}
,\frac{1}{y_2} ,-\frac{1}{y_3}\right),\quad
Re\,f^3_i =\frac{e^{K/2} y_3}{2}\left(-\frac{1}{y_1},
-\frac{1}{y_2} ,\frac{1}{y_3}
\right).
\ea
Here and below $y_i$ are real tree-level moduli.
Also we find 
$$Im f^I_i =-\e\frac{a_I }{2y_i}e^{K/2}, \quad I,i =1,2,3.$$
Collecting the above expressions, we obtain the system
of three equations for three unknown functions $a_i$
\ba
\label{E94}
 a'_1 +\frac{4e^{U +K/2} y_1 }{r^2 }
\left[q_0 +q_1 y_2 y_3  -a_1 \h{P}^1 -a_a \h{Q}_a + 
\h{P}^0 (-a_1 y_2 y_3 + a_2 y_1 y_3  +a_3 y_1 y_2 )\right]=0\nonumber \\
 a'_2 +\frac{4e^{U+ K/2} y_2 }{r^2 }
\left[q_0 -q_1 y_2 y_3  -a_1 \h{P}^1 -a_a \h{Q}_a +
\h{P}^0 (a_1 y_2 y_3 - a_2 y_1 y_3  +a_3 y_1 y_2 )\right]=0\nonumber \\
 a'_3 +\frac{4e^{U+ K/2} y_3 }{r^2 }
\left[q_0 -q_1 y_2 y_3 - a_1 \h{P}^1 -a_a \h{Q}_a +
\h{P}^0 (a_1 y_2 y_3 + a_2 y_1 y_3  -a_3 y_1 y_2 )\right]=0.
\ea

Substituting the explicit tree-level expressions for the moduli and metric
and using definitions (\ref{E47}), we obtain solution of the
Eqs. (\ref{E94})
\ba
&{}&a_1 = \frac{r}{2(P+r)}\left[c_1 +\frac{q^{(2)}}{r} +c_+ ln\frac{Q+r}{P+r} -
\frac{q^{(1)}}{Q}ln\frac{r}{r+Q} +\frac{q^{(1)}}{P}ln\frac{r}{r+P}\right],
\\\nonumber
&{}&b_2 -b_3 = \frac{c_- r}{r+P},\\\nonumber
&{}&b_2 +b_3 = \frac{c_+ r +q^{(1)}}{r+Q}.
\ea
Here 
$$b_a =\frac{a_a}{y_a} \qquad q_0 -q_1 y_2 y_3 \equiv
\frac{q^{(1)} e^{\g_0}}{\sqrt{8}},\qquad
q_0 +q_1 y_2 y_3 \equiv
\frac{q^{(2)} e^{\g_0}}{\sqrt{8}}.$$

As it was discussed in Sect.3, to have a static metric, we must take a
solution for which $Im\, T^-_{0n} =0$ or $Re \,S_{0n}=0$ which reads
\be
\label{E95}
2Pa_1 +(b_2 +b_3 )(P y_1 + Q) - q^{(2)} =0. 
\ee
Logarithmic terms should vanish separately, thus
$$q^{(1)} =c_+ =0.
$$
Thus, solution which retains the metric static is
\ba
\label{E96}
&{}&a_2 e^{-\g_0 -\s_0} = -a_3 e^{-\g_0 +\s_0}
=c_-\frac{r}{r+P}, \\\nonumber
&{}&a_1 =\frac{q^{(2)}}{2P}.
\ea
Because the axion $a_1$ is defined up to a constant, this means that
we can consistently set  $a_1 =0$.
Condition ${q^{(1)}}=0$ reads $q_0 =q_1 e^{2\g_0}$.

General solution contains three arbitrary constants which
 can be adjusted so that the asymptotic ("physical")
charges of the electric fields (\ref{B13}) and symplectic invariant
$Im\,T_{0n}^-$ responsible for non-stationarity of the metric vanish. 
\section{Discussion}
In this paper we obtained the loop-corrected dyonic solution  by solving the
 system of the loop-corrected Maxwell and Killing spinor equations.
Loop-corrected equations of motion were derived in perturbative approach
in the first order in string coupling constant from the loop-corrected
$N=2$ supersymmetric effective action which has only one-loop perturbative
corrections.

Except for a general theoretical interest of the study which
involves a number of subtle points such as an account of ambiguities of the
loop corrections to the prepotential and gauge couplings, possible
modifications of the symplectic transformation connecting sections
with the prepotential and associated with the heterotic string
compactification, etc., explicit form of the loop corrections provide
bounds on the applicability of the usually discussed tree-level solutions.

In the case of dyonic black hole with electric charges subject
to the condition $(Q_2 Q_3
)^{1/2}>\e (P^0 P^1)^{1/2} |h|/{TU}$ the loop-corrected solution
(\ref{M1}) is valid for all $r$. For magnetic black hole
with the charges $P^0$ and $P^1$ the loop-corrected solution (\ref{M2}) is
valid for $r>\e (P^0 P^1)^{1/2} |h|/{TU}$ and condition of the
applicability of the perturbative expansion is $r>\e (P^0 P^1)^{1/2}$.
 
Although the tree-level dyonic black hole is static
and with real moduli, at the one-loop  level the appearance
of  extra electric charges and imaginary parts of the moduli (axions)
as well as non-stationarity parameters in the metric
of the first order in string coupling is possible.
The asymptotic (physical) charges of the extra
electric field strengths depend on the axions, and by a suitable choice of
free constants can be set to zero; extra magnetic charges with the
accuracy of the first order
in string coupling  do not enter the equations of motion and can
be consistently set to zero.
Solving the equations for the axions, we show
that, as at the tree level,
 there exists a stationary loop-corrected solution with four
charges, but with the non-vanishing axions.

In perturbative approach, all the expressions which are of the first order
in string coupling and depend on the moduli are calculated by substituting
the tree-level moduli. Considerable simplifications are achieved for constant
tree-level moduli $T$ and $U$.
The results for dyonic black holes with four charges and full
coordinate dependence of the moduli will be published elsewhere.
Although the tree-level moduli are constant, the loop corrections to the
moduli and the metric have full coordinate dependence.
 
Except for the special values of the charges $P$ and $Q$, the moduli are away
from the enhanced symmetry points, where second derivatives
of the prepotential have logarithmic singularities \cite{wikalu,afgnt,hm}.

The expressions for the loop corrections are valid for all $r$ for which 
 the perturbation expansion in string coupling is valid. For the
dyonic solution, the sufficient condition is $\e\frac{P}{Q} <1$. 
In magnetic case,
both the tree-level and the loop-corrected solutions can be used in the
range $\frac{r}{P}>\e V$.
Perturbative corrections to dyonic and purely electric black holes,
 have finite limits $r\rightarrow 0$, in magnetic case the limit
 $r\rightarrow 0$ is singular.  

In this paper we considered the $N=2$ supersymmetric $STU$ model as
stemming from the suitably compactified heterotic string theory.
However, different embeddings of the $STU$ model in the underlying
string theory are possible \cite{duff,cvyo,bh2}. In  papers 
\cite{ch,bt} and refs. therein general classical $4D$
BPS-saturated generating black-hole solution preserving $1/8$ of $N=8$
$4D$ supergravity was constructed, 
and its $NS-NS$ and $R-R$ embeddings in the type
IIA and IIB theories were studied. 
The U-duality orbit of a solution of the $N=2$ truncation of  $N=8$
supergravity with $1/2$ supersymmetry unbroken connects solutions of type
II and heterotic theories with the same amount of supersymmetry unbroken.
In different embeddings of the $STU$ model in the underlying theory
the moduli and $4D$ vector fields stem from different fields
of the underlying higher-dimensional theory.
Heterotic embedding, while less suitable for calculation of
the microscopic entropy, is more natural for the study
of string-loop corrections. In this case, dilaton  is a natural
string-loop expansion parameter, 
and the axion-dilaton parametrize a separate factor
of the scalar manifold of the $STU$ model.

String-loop corrections to the classical charged black
holes as solutions of the effective $N=2$ $STU$ model the with string-loop
corrected prepotential was studied also in
papers~\cite{b6,bls,bgl}.  However, the two approaches are rather
different. In these papers, the loop-corrected metric
was calculated starting from the prepotential of the type II theory 
compactified on a Calabi-Yau threefold, which is dual to heterotic string
compactified on $K3\times T^2$. The prepotential of type II theory is of
purely classical origin, but has the same structure as the heterotic one
(the prepotentials are identical under a suitable identification
of the moduli in both theories). Perturbative solution was obtained 
under the assumption that there exists a "small" modulus on the type II side, 
and it is 
possible to expand the loop correction to the prepotential with
respect to the ratios of small to large moduli. As we discussed above, 
within the framework
of the heterotic string  compactification on $K3\times T^2$, the natural 
loop-expansion parameter is
expressed via the dilaton. However, this modulus does not enter the loop
correction to the prepotential. The remaining  moduli, which are connected
to the metric and antisymmetric tensor 
of the compact two-torus, for special configurations  
may have parametric smallness, but not the functional one connected
with dependence on $r$. Moreover, as we have argued above, to study
the loop-corrected solution it is important to take into account  
corrections to the gauge couplings.

Electrically charged black holes in dilatonic Einstein-Maxwell gravity
with constant threshold corrections were considered in \cite{chan}.
However, the model lagrangian of this paper is different
from the loop-corrected one of the present study, and hence the results are
incomparable.

Solutions of the equations of motion derived from $4D$ heterotic
string effective action with $N=1$ supersymmetry with one-loop perturbative 
corrections (note that
in this case there are contributions from all loops) and
also non-perturbative corrections included were studied in \cite{t3}. However,
in contrast to the present paper, only solutions with the flat space-time as
well as with other simplifications were considered. 

\vspace{.5cm}
{\large \bf Acknowledgments}

I would like to thank the members of the theoretical seminars at
the Skobeltsyn Nuclear Physics Institute and  Lebedev Physical Institute for
 discussions, M.Bertolini and M.Trigiante for useful e-mail correspondence
and K. Behrndt for a remark.
This work was partially supported by the RFFR grant No 02-02-16444.


\end{document}